\documentclass[11pt]{article}
\def\be{\begin{equation}}
\def\ee{\end{equation}}
\def\la{\label}
\def\bea{\begin{eqnarray}}
\def\eea{\end{eqnarray}}
\def\non{\nonumber}
\def\ci{\cite}
\def\la{\label}
\def\bib{\bibitem}

\def\lm{\lambda}
\def\Lm{\Lambda}
\def\le{\left}
\def\ri{\right}

\def\gm{\gamma}
\def\gmp{\gamma_\phi}

\def\Omp{\Omega_\phi}
\def\Ompi{\Omega_{\phi i}}
\def\Ompsc{\Omega_{\phi sc}}
\def\Om{\Omega}
\def\wp{w_\phi}
\def\wpsc{w_{\phi sc}}
\def\Ompo{\Omega_{\phi o}}
\def\wpo{w_{\phi o}}
\def\weff{w_{eff}}

\def\fr{\frac}
\def\pp{\partial}
\def\raw{\rightarrow}

\usepackage{graphicx}

\begin{document}

\begin{flushright}
  astro-ph/0110460 \\

\end{flushright}

\vspace{15mm}

\begin{center}
   {\Large \bf Quintessence Restrictions on Negative Power \\
and Condensate Potentials}
\end{center}

\vspace*{0.7cm}

\begin{center}
{\bf A. de la Macorra\footnote{e-mail:
macorra@fisica.unam.mx} and C. Stephan-Otto
\footnote{e-mail: stephan@servidor.unam.mx}}
\end{center}

\vspace*{0.1cm}

\begin{center}
\begin{tabular}{c}
{\small Instituto de F\'{\i}sica, UNAM}\\ {\small Apdo. Postal
20-364, 01000  M\'exico D.F., M\'exico}\\
\end{tabular}
\end{center}

\vspace{1 cm}

\begin{center}
{\bf ABSTRACT}
\end{center}
\small{We study the cosmological evolution of scalar fields that
arise from a phase transition at some energy scale $\Lm_c$. We
focus on negative power potentials given by
$V=c\Lm_c^{4+n}\phi^{-n}$ and restrict the cosmological viable values
of $\Lm_c$ and $n$. We make a complete analysis of $V$ and impose
$SN1a$ conditions on the different cosmological parameters. The
cosmological observations ruled out models where the scalar field
has reached its attractor solution. For models where this is not
the case, the analytic approximated solutions are not good enough
to determine whether a specific model is phenomenologically viable
or not and the full differential equations must be numerically
solved. The results are not fine tuned since a change of $45\%$
on the initial conditions does not spoil the final results.
We also determine  the values of $N_c, N_f$ that give a
condensation scale $\Lm_c$ consistent with gauge coupling
unification, leaving only four models
that satisfy unification  and SN1a constraints. }

%\vspace*{1cm}

\noindent \rule[.1in]{14.5cm}{.002in}

\thispagestyle{empty}

\setcounter{page}{0} \vfill\eject

\section{INTRODUCTION}

Recent cosmological results given by the superonovae project SN1a
\ci{SN1a} and the Maxima and Boomerang \ci{CMBR} observations
have lead to conclude that the universe is flat and it is
expanding with an accelerating velocity.  This conclusions show
that the universe is now dominated by a energy density with
negative pressure with $\Om_{phi}=0.7 \pm .1$ and $ w_\phi < -2/3$
\ci{w} . This energy is generically called the cosmological
constant. An interesting parametrization of this energy density is
in terms of a scalar field with gravitationally interaction only
called quintessence \ci{tracker}. The evolution of scalar field has
been widely studied and some general approach con be found in
\ci{generic}, \ci{mio.scalar}. The evolution of the scalar field
$\phi$ depends on the functional form of its potential $V$ and a
late time accelerating universe constraints the form of the
potential \ci{mio.scalar}.

In this paper we will concentrate on negative power potentials
because they lead to an acceptable phenomenology and because they
are naturally obtained from gauge group  dynamics. Negative power
potential have been extensively studied \ci{tracker}-\ci{mio.scalar}
 first by \ci{1/q} and
then as tracker fields by \ci{tracker}. Steinhardt et al.
\ci{tracker} showed that scalar field with a negative power potential
$V=c \phi^{-n}$  with $n
> 5$ has already reached its tracker solution but are not
cosmologically acceptable because they have $w_{\phi } > -0.52$.
 However, if the scalar field has not reached its
tracker solution by today, we will show that the models may
lead to an acceptable phenomenology and the final results depend
on the initial conditions and on the value of $n$. Contrary to the
tracker models, no analytic solution is good enough to determine
the value of
  $w_{\phi o}$ (from now on the subscript $o$ will
refer to present days values) and it is  sensitive to the whole
dynamics. We will solve numerically the differential equations and
we will constrain the values $n$ of cosmologically viable models,
including the big bang nucleosynthesis (NS) constraints \ci{NS}.
We will also give approximated analytically solutions.

Tracker solutions are widely favored because they do not have a
fine tuning problem of initial conditions. But even more, they are
independent of the initial conditions since the range of the
initial conditions can vary up to 100 orders of magnitude. The
models with $n < 5$ do depend  on the initial conditions but it is
important to remark that they {\it do not} have a fine tuning
problem. The initial conditions can vary up to 45\%, the
solutions are still fine and  the values of the initial conditions
are completely "natural", i.e. they are of the same order of
magnitude as the other relevant cosmological parameters. So, to
conclude, one thing is to have a model with no dependence on the
initial conditions and another is to have a fine tuning problem.
"Natural" models in physics should not have fine tuning problems
but they do in general depend on the initial values as it is the
case for our models. However, for any initial conditions we
will end up with  $-1 \leq \wpo \leq -2/(2+n)=w_{tr}$, where
$n$ gives the inverse power and $w_{tr}$ is the tracker value.

Negative power potentials \ci{bine}-\ci{chris1} can
be obtained using Affleck et al. ("ADS") superpotential
\ci{Affleck}. The condensation scale $\Lm_c$ of the gauge group
$SU(N_c)$ can be determined from a high energy scale ($\Lm$) using
renormalization group equation in terms of $N_c, N_f$ and it is
then naturally to ask if it is possible to have a common gauge
coupling unification with the Standard Model ("S.M.") gauge groups
\ci{chris1}. We will show that this is indeed possible and we will
give the values of $N_c, N_f$ where gauge coupling unification is
achieved.

The cosmological picture in the case of gauge coupling unification
is very pleasing. We would assume gauge coupling unification at a
scale $\Lm$ for all gauge groups (as predicted by string theory)
and then let all fields evolve. At the beginning all fields,
"S.M." and those from the $SU(N_c)$ gauge group, are massless and
redshift as radiation until  we reach the condensation scale
$\Lm_c$. Below this scale  the fields of $SU(N_c)$ group will
dynamically condense and we use ADS potential to study its
cosmologically evolution. Interesting enough, the relative energy
density of the $SU(N_c)$ group $\Omp$ drops quickly, independently
of the initial conditions, and it is close to zero for a long
period of time, which may include nucleosynthesis,  until very
recently (around one e-fold of inflation). The energy
density of the universe is at present time  dominated by the scalar field with
$\Omp \simeq 0.7$ and a negative pressure $w_\phi <-2/3$ leading
to an accelerating universe \ci{w}.

The paper is organized as follows. In section \ref{beha} we give
the general frame work to derive the scalar potential for $\phi$
from non-abelian gauge dynamics using ADS potential. In section
\ref{evol} we analyze the cosmological evolution of $\phi$ and we
concentrate  in the non attractor regime. We derive analytic
formulas for $\wpo$ and $\phi_o$ as a function of the initial
conditions and on $n$ and we discuss in detail the possible
choices of initial conditions and show that the models do not
have a fine tuning problem. In section
\ref{sec12} we constrain the values of $n$ in order to have
$\wpo<-2/3$ while in section \ref{unif} we comment on the
possibility to have models with a gauge coupling constant unified
with the couplings of the standard model and we explicitly give
these models. In section \ref{examp} we give further examples.
Finally, we summarize and   conclude  in section
\ref{concl}.

\section{POTENTIALS OF THE FORM $V=c\Lm_c^{4+n}\phi^{-n}$}\la{beha}

In this work we study quintessence field (scalar) $\phi$  with
negative power potentials that arise from a phase transition at
some stage of the evolution of the universe. The energy scale of
the phase transition is given by $\Lm_c$ and the initial value of
the scalar field $\phi$ is naturally given by $\phi_i=\Lm_c$ since
it is the relevant scale for the transition. The potential we will
consider is of the type
\be
V=c^2 \Lm_c^{4+n}\phi^{-n}
\label{pot}
\ee
with $c$ a constant (we
will comment on the value of $c$ in section \ref{sec12}) and has a
runaway behaviour.

This class of models would have a vanishing potential $V\equiv 0$
for energy scales above $\Lm_c$ since the phase transition has not
taken place yet and there is no $\phi$ field. At the
phase transition energy scale $\Lm_c$ a potential
  $V(\Lm_c)\simeq\Lm_c^4$ and a field $\phi(\Lm_c)=\Lm_c$  are  generated.
Below $\Lm_c$, the $\phi$ field becomes dynamically and it
evolves to its minimum. The cosmological evolution depends on the
functional form of $V$ and for eq.(\ref{pot}) with $n>0$ we
expect $\phi$ to roll down its potential. This class of
potential has been chosen because they can be obtained from a
phase transition of
non-abelian gauge dynamics (see subsection \ref{ads})
\ci{Affleck} and because they lead to a quintessence interpretation
of the $\phi$ field. However, if other physical process leads
also to inverse power scalar potential the cosmological evolution
studied in section \ref{evol} and the conclusions remain valid.

The energy scale $\Lm_c$ is expected to be considerably smaller
than the reduced Planck Mass, $m_{Pl}$, so the initial value
$\fr{\phi_i}{m_{Pl}}= \frac{\Lm_c}{m_{Pl}}$ is much smaller than 1
and this has interesting consequences for the cosmological
evolution of $\phi$ (we will set from now on the reduced Planck
mass to one $m_{Pl}^2=G/8\pi\equiv 1$).

The normalization of the field is important and
 we will consider, for simplicity,  the $\phi$ field to be
canonically normalized, $L_k=(K_\phi^\phi)^{-1}|\pp_\mu \phi|^2$
with $K_\phi^\phi=1$ and $K_\phi\equiv\pp K/\pp \phi$. However,
the complete Kahler potential $K$ is in general not known. The
canonically normalized field $\phi'$ can be defined by
$\phi'=g(\phi,\bar{\phi})\phi$ with the function $g$ given from
solving the differential equation $K_\phi^\phi=(g+\phi
g_\phi+\bar{\phi}g_{\bar{\phi}})^2$. For $\phi \ll 1$ we do not
expect any large contributions to the kinetic term but for $\phi
\sim 1$ the Kahler potential could give a significant contribution
and  could spoil the runaway and quintessence behaviour of $\phi$.
In order to see this, we can expand the Kahler potential as a
series power $K=|\phi|^2+\Sigma_{i}a_i |\phi|^{2i}/2i$ with $a_i$
some constants of order one and to be determined by the specific
model. If we approximate, for simplicity, the canonically
normalized field $\phi'$ by $\phi'=(K_\phi^\phi)^{1/2}\phi$. The
potential in eq.(\ref{pot}) would then be given by
\bea \la{v}
V&=&(K_\phi^\phi)^{-1}|W_\phi|^2=
c^2\Lm_c^{4+n}\phi^{-n}(K_\phi^\phi)^{-1}
\nonumber\\
&=&c^2 \Lm_c^{4+n}\phi'^{-n} (K_\phi^\phi)^{\fr{n}{2}-1}
 \eea
 For $n<2$ the
exponent term of $K_\phi^\phi$ in eq.(\ref{v})  is negative so it
will not spoil the runaway behaviour of $\phi$ but for $n>2$ the
extra terms could stabilize the potential. In the lack of a better
understanding of $K$ we will work with canonically normalized
fields but we should keep in mind that for $n<2$ the results are
robust while for $n>2$ the contribution from the Kahler potential
could spoil our results and must be determined.

\subsection{ADS Potential}\la{ads}

The potential in eq.(\ref{pot}) can be obtained from the
non-perturbative dynamics of a non-abelian asymptotically free
gauge group $SU(N_c)$
with $N_f$ chiral + antichiral fields $Q$  in N=1 supersymmetric
theory. At energy scales much larger than the condensation scale the gauge
coupling constant is small and the $Q$ fields are free elementary
fields. As the universe expands and cools down, the energy of
the elementary fields $Q$ becomes smaller while  the gauge coupling
constant grows. When the gauge coupling constant has the critical
value to condense the $Q$ fields, then all the elementary fields
will no longer be free and they will form "mesons" and "baryons",
as in QCD. This effect takes place at the condensation scale
$\Lm_c$ and below this scale the correct description of the
dynamics of the non-abelian gauge group is in terms of the
condensates $\phi$. In order to study the dynamics of these
fields $\phi$ we use the ADS superpotential, which is  exact
(i.e. it does not receive  radiative or non-perturbative
contributions) and  it is given by $
 W(\phi)=(N_c-N_f)\left(\fr{\Lm_c^{3N_c-N_f}}{det <Q\tilde
Q>}\right)^{1/(N_c-N_f)}$  \cite{Affleck}. In terms of the gauge
singlet  combination of chiral and antichiral bilinear term
$\phi=<Q\tilde Q>$  the globally supersymmetric scalar potential
is given by eq.(\ref{v})  with $c=2N_f$, $det<Q \tilde
Q>=\Pi_{j=1}^{N_f}\phi_j^2$ and $ n=2+\fr{4N_f}{N_c-N_f}$
\ci{bine}-\ci{chris1}.

If we wish to study models with $0<n<2$, which are cosmologically
favored as we will see in section \ref{sec12}, we need to consider the
possibility that not all $N_f$ condensates $\phi_i$ become
dynamically  but only a fraction $\nu$ are (with $N_f\geq\nu\geq
1$) and we also need $N_f > N_c$. It is important to point out
that even though it has been argued that for $N_f>N_c$ there is no
non-perturbative superpotential $W$ generated \ci{Affleck} this is
not always the case \ci{ax.asy}. The possibility of having $\nu
\neq N_f$ can be done with gauge group with unmatching number of
chiral and anti-chiral fields or if some of the chiral fields are
also charged under another gauge group. In this later case we have
$c=2\nu, n=2+\fr{4\nu}{N_c-N_f}$ and $N_f-\nu$ condensates fixed
at their v.e.v. $<Q \tilde Q>=\Lambda_c^2$ \ci{chris1}.

\section{Evolution of $\phi$}\la{evol}

We will now determine the cosmological evolution of a scalar field
 $\phi$ with  an
inverse power potential regardless from what physical process it
arose. We will concentrate on potentials that have no reached the
tracker solution yet since they can give  the correct value of $\wpo$
and we will give approximately analytic solutions to $\wpo$ and
$\phi_o$.

The cosmological evolution of $\phi$ with an arbitrary potential
$V(\phi)$ can be determined from a system of differential
equations describing a spatially flat Friedmann--Robertson--Walker
universe in the presence of a barotropic fluid energy density
$\rho_{\gm}$ that can be either radiation or matter, are
\bea\la{eqFRW}
\dot H &=& -\frac{1}{2}(\rho_\gamma+ p_\gamma+\dot \phi^2),\non \\
\dot \rho &=& -3H(\rho+p),\\
\ddot \phi &=& -3H \dot \phi-\frac{d V(\phi)}{d \phi},\non
 \eea
where $H$ is the Hubble parameter, $\dot \phi = d\phi/d t$, $\rho$
($p$) is the total energy density (pressure).  We use
the change of variables $x \equiv \frac{\dot
\phi}{\sqrt{6} H}$ and $y \equiv \frac{\sqrt{V}}{\sqrt{3} H }$
   and  equations (\ref{eqFRW}) take the following form
  \cite{liddle,mio.scalar}:
 \bea  \la{eqFRW2}
x_N&=& -3 x +\sqrt {3 \over 2} \lambda\,  y^2 + {3 \over 2} x
[2x^2 + \gm_{\gm} (1 - x^2- y^2)] \non\\
 y_N&=& - \sqrt {3 \over
2} \lambda \, x\, y + {3 \over 2} y [2x^2 + \gm_{\gm} (1 - x^2
-y^2)]\\
H_N&=& -{3 \over 2} H [2x^2 + \gm_{\gm} (1 - x^2 - y^2)]
\non
 \eea
where $N$ is the logarithm of the scale factor $a$, $N\equiv
ln(a)$; $f_N\equiv d f/d N$ for $f=x,y,H$; $\gm_\gm= 1+ w_\gm $
and $\lambda(N)\equiv -V'/V$ with $V'= dV/d\phi$. In terms of $x,
y$ the energy density parameter is $\Om_{\phi}=x^2+y^2$ while the
equation of state parameter is given by $w_{\phi}\equiv
\rho_{\phi}/p_{\phi}=\frac{x^2-y^2}{x^2+y^2}$.

The Friedmann or constraint equation for a flat universe
$\Omega_{\gm}+\Om_{\phi}=1$ must supplement equations
(\ref{eqFRW2}) which are valid for any scalar potential as long as
the interaction between the scalar field and matter or radiation
is gravitational only. This set of differential equations is
non-linear and for most cases has no analytical solutions. A
general analysis for arbitrary potentials is performed in
\cite{mio.scalar}, the conclusion there is that all model
dependence falls on two quantities: $\lambda(N)$ and the constant
parameter $\gm_\gm$. In the particular case given by $V \propto
1/\phi^n$ we find $\lm \raw 0$ in the asymptotic limit. If we
think the scalar field appears well after Planck times we have
$\lm_i=n\,m_{Pl}/\Lm_c \gg 1$  (the subscript $i$ corresponds to
the initial value of a quantity). An interesting general property
of these models is the presence of a many e-folds scaling period
in which $\lm$ is practically a constant and $\Om_{\phi} \ll 1$.
Figure \ref{lambda} shows the rapid arrival and long permanence of
this parameter to its constant value, together with the final
decay to zero. On this last regime we have $\lm \raw 0$ that
implies $\frac{x_N}{x}<0$ and $\frac{y_N}{y}>0$ \cite{mio.scalar},
leaving us with $\Om_{\phi} \equiv x^2+y^2 \raw 1$ and $\wpo
\equiv \frac{x^2-y^2}{x^2+y^2} \raw -1$, which are in accordance
with a universe dominated by a quintessence field whose equation
of state parameter agrees with positively accelerated expansion.
\begin{figure}[htp!]
\begin{center}
\includegraphics[width=8cm]{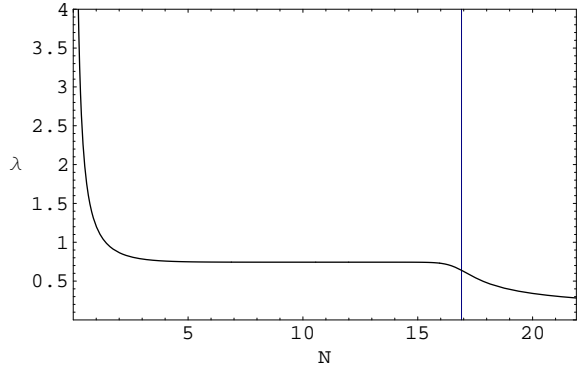}
\end{center}
\caption{\small{Evolution of $\lambda$ for $n=1$. The
vertical line marks the time at $\Ompo=0.7$  with
$N_{total}=17.04$.}}\la{lambda}
\end{figure}
The development of $\Omp$ can be in agreement with the restriction of
the nucleosynthesis stage $\Omp(\mathrm{NS})<0.1$ \cite{NS} as
well as with the observational result $\Ompo = 0.7$ (the subscript
$o$ refers to present day quantities). This can be observed in
Figure \ref{Oomega}, together with the evolution of $\wp$ which
fulfills the condition $\wpo < -2/3$ \ci{w}.

 The analysis of inverse power potentials has been extensively
 studied \ci{tracker}-\ci{mio.scalar}. However, the analysis has not
  been specific enough
 to determine their viability to describe the late evolution of
 our universe. In Steinhardt et al. \ci{tracker} the scalar field was required
 to track before present day and this imposes a constraint on $n$
 to be larger than 5 and thus ruling this models out since they
 have $\wpo > -0.52$ in contradiction to SN1a data. The models we
 will concentrate on are, therefore, models with $n < 5$ where $\phi$ has not
 reached its tracker value.

For future reference  we give now the  scaling value
of\footnote{Our value of $\phi_{sc}$ differs in the case of $\Ompi
>1/2$ by a factor of $1/\sqrt{2}$ from \ci{tracker} and they use $M_{Pl}=1$
 instead of $m_p=M_{Pl}/\sqrt{8\pi}=1$ as we do} $\phi$ \ci{tracker}
 \bea \la{psc}
 \phi_{sc}&=&\phi_i+\sqrt{6\Ompi} \hspace{4cm}
  for \;\; \Ompi < 1/2
  \non\\
  \phi_{sc}&=&\phi_i + \sqrt{6}\le(\fr{1}{\sqrt{2}}+\fr{1}{2}Log[\fr{\Ompi}{1-\Ompi}]\ri)
\hspace{1cm} for \;\;\Ompi > 1/2
\eea
The scaling value only depends on the initial conditions $\Ompi$, it is
 independent on $\Lm_c, H_i$,  since $\phi_i \ll 1$. The tracker value of $w$ is
given by \ci{tracker}
 \be \la{wtr}
 w_{tr}=-1+\fr{n}{2+n}(w_\gm +1)
\ee
and it is an attractor solution valid for large $n$, when $\phi$ is
 already tracking. In the tracker limit \ci{tracker}, i.e. $n=5$,
 from eq.(\ref{wtr})
one has  $w_{tr}=-0.28$ but the value obtained numerically is
 only $w_o=-0.52$ for $\Ompi > 0.25$. For smaller $n$ the
 discrepancy is even worse since the
 scalar field has not reached its tracker value, obtained from
eqs.(\ref{wtr}) and (\ref{wo}),
\be
\phi_{tr}=\sqrt{\fr{n(2+n)\Ompo}{3}},
\ee
which is larger than $\phi_{sc}$ if $\Ompi <  \fr{n(2+n)}{18}\Ompo $.

A semi-analytic approach is useful to study some properties of the
differential equation system given by eqs.(\ref{eqFRW2}). To do this we
initially consider only the terms that are proportional to $\lm$,
since $\lm \gg 1$, then we follow the evolution of $x$, $y$ and $H$ so every
period has a characteristic set of simplified differential
equations. The parameter $\Omp$ is adequate to divide the process
into four periods, the first one being a short lapse in which
$\Omp=constant$, easier to recognize in Figure \ref{x-y}; the
second is defined from the fall of this parameter to negligible
values; the third is the so-called scaling period and in the
fourth $\Omp$ is again considerable, eventually reaching the value
$0.7$.
\begin{figure}[htp!]
\begin{center}
\includegraphics[width=8cm]{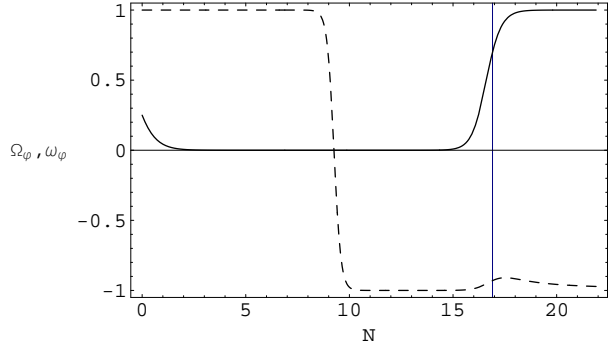}
\end{center}
\caption{\small{Evolution of $\Omp$ (solid curve) and $\wp$
(dashed curve) for $n=1$.  The
vertical line marks the time at $\Ompo=0.7$.}}\la{Oomega}
\end{figure}

The phase plane $x-y$ provides an illustrative approach, useful
for the analysis. We see, from Figure \ref{x-y}, that the system
follows, at first, a circular path with $\equiv x^2+y^2 =\Ompi$
constant and ends up with $x^2\simeq \Ompi \gg y^2$. Then, $x$ and
$y$ decrease to negligible values, this situation prevails all
along the scaling period. Finally, a growth in both parameters
causes $\Omp$ to reach what we set as the final value $0.7$,
preferred by observational results. From the restriction over the
equation of state parameter, $\wp < -2/3$ and the observational
range for $\Ompo=0.7 \pm 0.1$, we can define a region limited by
the expressions: $y^2=\frac{1-\wp}{1+\wp}x^2$ with $\wp=-2/3$ and
$y^2=\Ompo - x^2$ with $0.6 \leq \Ompo \leq 0.8$.
\begin{figure}[htp!]
\begin{center}
\includegraphics[width=9cm]{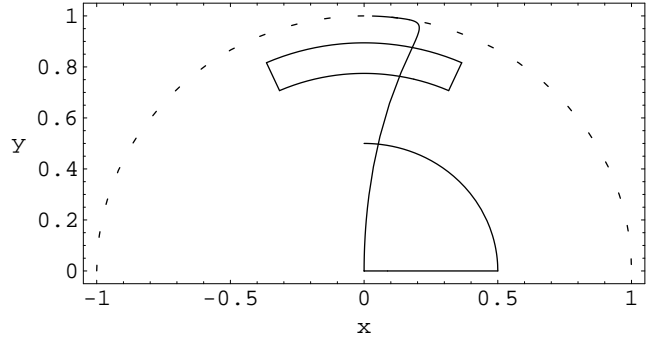}
\end{center}
\caption{\small{Phase plane $x-y$ for $n=1$. Starting point is
$(0,0.5)$. The region defined by $\wpo<-2/3$ and $\Ompo = 0.7 \pm
0.1$ is shown.}}\la{x-y}
\end{figure}

The minimal value $y_{min}$ of $y$ after its initial steep descent
is given from eq.(\ref{eqFRW2}) with $y_N=0, x^2\simeq \Ompi$ and
$\lm x \sqrt{3/2}=-H_N/H=3\gm_\gm/2$ by
\bea\label{ymin}
 y_{min}&=&y_i \left(\fr{\Lm_c}{\phi_{min}}
 \right)^{\fr{n}{2}}
 \non\\
\phi_{min}&=&\fr{n}{4}\sqrt{6\Ompi}
 \eea
 and we have approximated  $H_{min}\simeq H_i$ in eq.(\ref{ymin}).
 Shortly after $y$ reaches its minimum value the
  scaling period begins. In
this period we neglect the term proportional to $\lm$ in
eqs.(\ref{eqFRW2}) to find:
\be \la{yNHN}
\frac{y_N}{y}=-\frac{H_N}{N}
\ee
which leads to $yH=H_{min} y_{min}\simeq H_i y_{min}$.  Notice that the dependence
of $\lm$ from $y$ and $H$ is given by $\lm = A\ (y H)^{2/n}$ with
$A$ a constant, therefore from eq.(\ref{yNHN}) we have
$\lm=constant$ (i.e. $\phi=constant$) during  all of the scaling period, this holds for
any $n$. Furthermore, we may neglect squared terms on $x$ and $y$
in the third equation of system (\ref{eqFRW2}), since they are
small, to get the expressions
\bea \la{Hscal}
H&=&H_{i} e^{-\frac{3}{2}\gm_\gm N}
\nonumber\\
y&=&y_{min} e^{\frac{3}{2}\gm_\gm N}
\eea
The quantity $y$ has an increasing exponencial form for almost all
of the process, so the duration of this regime can be seen as the
total time (see Figure \ref{xyO}). Now, in order to calculate the
number of  e-folds from  the initial value to present day we
consider equation (\ref{Hscal}) to end up with:
\be  \la{Ntotal}
N_{total}=\frac{2}{3} \ln[\fr{y_o}{y_{min}}]
\ee
with $y_{min}$ given by eq.(\ref{ymin}) and $y_o\simeq 0.8$ (to
have $\Ompo=.7, \wpo<-2/3$. The evolution
of $\ln(x)$, $\ln(y)$ and $\ln(H)$ as a function of  $N$ is seen on Figure
(\ref{xyO}).
\begin{figure}[htp!]
\begin{center}
\includegraphics[width=9cm]{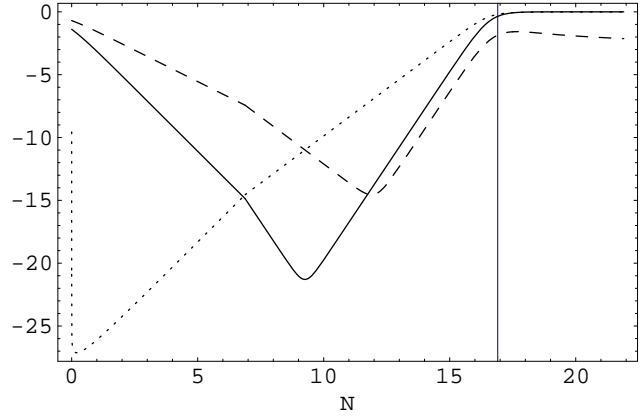}
\end{center}
\caption{\small{Evolution for $n=1$ of $ln(x)$, $ln(y)$ and
$ln(\Omp)$ (dashed, dotted and solid lines,
respectively).  The
vertical line marks the time at $\Ompo=0.7$.}}\la{xyO}
\end{figure}

If we consider eqs.(\ref{Hscal}) and assume that the end of the
scaling period is very close to today we get an approximated
equation $y_o H_o\simeq H_{min}y_{min}=H_iy_i(\phi_i/\phi_{min})$.
This, together with eq.(\ref{ymin}) and the definition of
$y^2_i=c^2\Lm_c^4/3H^2_i$ establishes an expression for $\Lm_c$,
the energy scale at which the scalar field appears, in terms of
$y_o$, $H_o$:
 \be \la{Lm}
\Lm_c=\left(\fr{3 y_o^2 H_o^2}{c^2} \right)^\frac{1}{4+n}
\phi_{min}^\frac{n}{4+n}
 \ee
 and $\phi_{min}=\fr{n}{4}\sqrt{6\Ompi}$.
The later expression is a semi-analytic calculation of the initial
energy scale of a specific model. Finally, the value $y_o$ is set
to be of the order of $0.8$ to satisfy simultaneously $\Ompo=0.7$
and condition $\wpo < -2/3$.

Of course, we could have guessed expression (\ref{Lm}) using the
definition of $y_o^2=\fr{c^2\Lm_c^{4+n}}{3H_o^2}\phi_o^{-n}$ to
give  \ci{bine}
\be\la{lmm}
\Lm_c=\left( \fr{3H_o^2y_o^2\phi_o^n}{c^2}\right)^{\fr{1}{4+n}}
\simeq H_o^{\fr{2}{4+n}}
\ee
and the last equality holds approximately since we expect to have
$y_o, \phi_o$ of the order of one.

Now, we wish to determine the value of $\wpo, y_o$ and $\phi_o$.
We use the differential equation
 for $\gmp=\wp+1$ and $\Omp$ \ci{ax.coinci}
 \bea\la{Dgm}
 (\gm_\phi)_N&=&3\gm_\phi(2-\gm_\phi)
 \le(\lm \sqrt{\fr{\Ompo}{3\gm_\phi}}-1\ri)
 \non\\
 (\Omp)_N&=&3(\gm_\gm-\gm_\phi)\Omp(1-\Omp).
 \eea
We see that $\gmp$ is extremized  at $\wp=\gmp-1=-1,1 $ and at
$\wp=-1+\fr{\lm^2\Omp}{3}=-1+\fr{n^2 \Omp}{3 \phi^2}$. We have
checked that the value of $\wp$ at the maximum evaluated at
$\phi_o$  is a very good approximation (within $5\%$ of the
numerical value)
 \be\la{wo}
 \wpo=-1+\fr{n^2 \Ompo}{3 \phi_o^2}.
 \ee
Eq. (\ref{wo}) should be compared with \ci{bine} which differs by
the factor of $\Ompo \simeq 0.7$. Of course, if we do not have the
exact value of $\phi_o$ eq.(\ref{wo}) is not very useful. In order
to determine $\phi_o$  we evolve eqs.(\ref{Dgm}) and
eq.(\ref{eqFRW2}) from present day values to the scaling regime
where $\wpsc\simeq -1$  and $x_{sc}^2 \ll y_{sc}^2\simeq \Ompsc $ with
 the  condition $\Ompo=0.7$. This evolution is model independent
if  $0\leq \wp+1 \ll 1$  \ci{ax.coinci} and
from the definition of $y^2$ we have
 \be \la{LM}
 \fr{c^2}{3}\Lm_c^{4+n}=y_o^2H_o^2\phi_o^n=y_{sc}^2H_{sc}^2\phi_{sc}^n
\ee
and we obtain
\be \la{yp}
y_o^2\phi^n_o=\fr{y_{sc}^2
H_{sc}^2\phi_{sc}^n}{H_o^2}=\phi^n_{sc} \Ompo
\ee
where we have used $H_N/H\simeq -3(1-\Omp)/2,
H_{sc}^2/H^2_o=(1-\Ompo)e^{3\Delta N}$ and
$ y^2_{sc} \simeq\Omp=\Ompo e^{-3\Delta
N}/(1-\Ompo +\Ompo e^{-3\Delta N}),\;\;\Delta N \gg 1$ \ci{ax.coinci}.
 Using eqs.(\ref{yp}), (\ref{wo})
and (\ref{psc}) we can solve easily for $\phi_o$ and/or $y_o$ in
terms of $\Ompo, n$ and $\Ompi$ (via $\phi_{sc}$),
 \be  \la{po}
  \phi_o^2 -\phi_{sc}^n\phi_o^{2-n}-\fr{n^2}{6}\Ompo=0
\ee
or equivalently
\be \la{yyo}
y_o^2\le(\fr{n^2\Ompo^2}{6(\Ompo-y_o^2)}\ri)^
{\fr{n}{2}}= \phi_{sc}^n\Ompo.
\ee
In order to analytically solve eqs.(\ref{po},\ref{yyo}) we need to fix the
value of $n$ and we can determine $\wpo$ by putting  the solution
of (\ref{po}) into eq.(\ref{wo}). Eq.(\ref{po}) can be rewritten as
$\phi_o=\phi_{sc}(1-n^2\Ompo/6\phi_o^2)^{-1/n}$
and we see that $\phi_o > \phi_{sc}$ and $\phi_o >
n\sqrt{\Ompo/6}$ and
  we see that $\phi_o$ is of the order of 1
 ($\Ompo\sim .7$)
regardless of the initial conditions. However, the exact value
does indeed depend on the initial conditions but for
any initial conditions we will have  $-1\leq \wpo \leq
w_{tr}$.

If $\gm_{\phi o}=n^2\Ompo/6\phi_o^2 \ll 1$ one
has $\phi_o \simeq \phi_{sc}$  and for the simple cases of $n=1,2$ and $4$
we can solve explicitly for $\phi_o$ and we find $\phi_o|_{n=1}=\phi_{sc}/2+\sqrt{9\phi_{sc}^2+6\Ompo}/6,
\;y^2_o|_{n=1}=\phi_{sc}(-3\phi_{sc}+\sqrt{9\phi_{sc}^2+6\Ompo},\;)$
$\phi_o|_{n=2}=\sqrt{\phi_{sc}^2+2\Ompo/3},\;y^2_o|_{n=2}=
3\phi_{sc}^2\Ompo/(3\phi_{sc}^2+2\Ompo)$
and $\phi_o|_{n=4}=\sqrt{4\Ompo/3
+\sqrt{9\phi_{sc}^2+16\Ompo}/3},\;$
$y^2_o|_{n=4}=\Ompo-8\Ompo^2/(4\Ompo+\sqrt{9\phi_{sc}^2+16\Ompo})$,
respectively. Notice that the value of
$\phi_o, \wpo$ at $\Ompo=0.7$ {\it does not} depend on $H_i$ or
$H_o$ and it only depends on $\Ompi$ (through $\phi_{sc}$) and
$n$.

We show in Figure
\ref{omw} how $\wpo$ varies for different initial conditions
$\Ompi$  with $n=18/7\simeq 2.57$
fixed. We see that for larger $\Ompi$ we end up with a smaller
$\wpo$ and this is a generic result as can be seen from
eqs.(\ref{po}),(\ref{psc}) and (\ref{wo}) since for larger $\Ompi$ one has a
larger $\phi_{sc}$,  $\phi_o$ and therefore a smaller $\wpo$. This
can be  seen also from Figure \ref{Oomega1} where for a small
$\Ompi$ a plateau arises in $\wp$ (the field $\phi$ has already reached its tracker
value by present day). From eqs.(\ref{wo}),(\ref{po}) we notice  that for
smaller $n$ one gets a smaller $\wpo$ (see Figure \ref{weff}).
\begin{figure}[p!]
\begin{center}
\includegraphics[width=9cm]{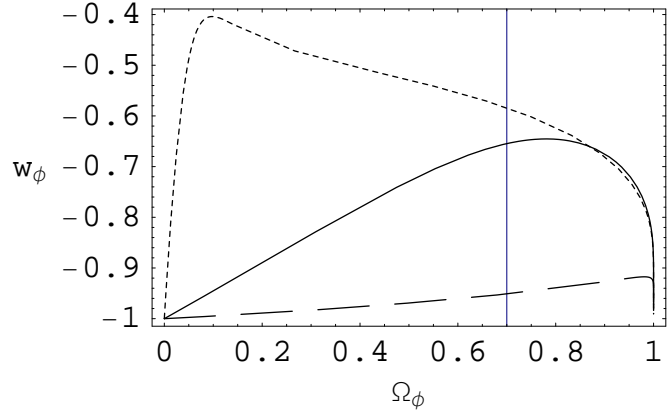}
\end{center}
\caption{\small{Variations on $\Ompi$ lead to different physical
situations given by $\wpo$. We have taken $\Ompi=0.9,0.25,10^{-10}$
dashed, solid
and dotted lines, respectively.  The
vertical line marks the time at $\Ompo=0.7$.}}\la{omw}
\end{figure}
\begin{figure}[p!]
\begin{center}
\includegraphics[width=9cm]{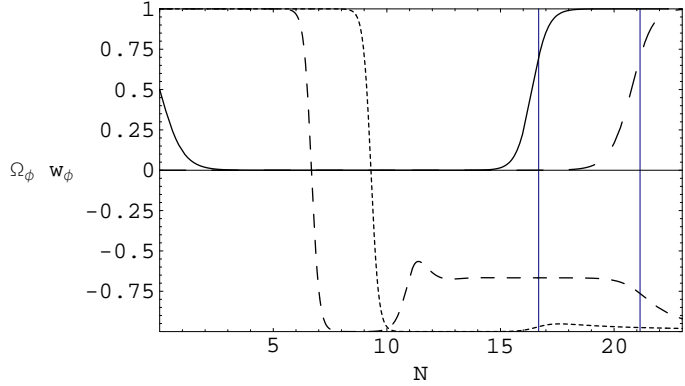}
\end{center}
\caption{\small{Evolution of $\Omp$ and $\wp$ for $n=1$. Parameter
$\Omp$ evolves in a similar way for $\Ompi=0.25$ (solid line) and
$\Ompi=10^{-10}$ (long-dashed line), varying only in the total
time, $N_{total}=16.7$ or $21.1$ in the previous order. A
discrepancy between the two cases at $\Ompo=.7$ is seen for
parameter $\wp$
because it has, for $\Ompi=0.25$  a local maximum with $\wpo=-0.93$
(dotted line) while for $\Ompi=
10^{-10}$ (short-dashed line) a
plateau appears with $\wpo=-0.76$. The
vertical lines mark the time at $\Ompo=0.7$. }}\la{Oomega1}
\end{figure}

\subsection{Initial Conditions}\la{init}

It is well known that inverse power law potential lead
to attractor solutions \ci{tracker} and are therefore independent of
the initial conditions (one has 100 orders of magnitud
range for $\Ompi$). However, as pointed out by Steinhardt et al,
this is only the case for $n>5$, for smaller values of $n$
the field $\phi$ has not necessarily reached its tracker value
at present time. However, even though in these cases the
present day quantities depend on the initial conditions
there is no fine tuning problem since we can vary the
initial conditions on a wide range of values (i.e. 45\%)
and the end results are still physical acceptable.

 The differential equations given in eqs.(\ref{eqFRW2})
 depend on values of $x,y, \lambda=n/\phi$ but they do not
 depend on the absolute value of H, i.e. we have
 the same evolution for $x,y,H$ as for $x,y,H'=k H$ where
 $k$ is an arbitrary constant. This scaling freedom allow
 us to set the normalization of $H$ as we wish and in particular
 to have  $\Ompo=0.7$ at $H=H_o$  for any values of the initial
 conditions $x_i, y_i$. This implies that we can have
 a quintessence model for arbitrary initial conditions.
 Once $H_o$ is fixed we get $\Lm_c\simeq H_o^{2/(4+n)}$
(eq.(\ref{lm})) and $H_i=c\Lm^2_c/y_i$ (from the definition of
$y_i$). One of the main difference with the tracker solution is
that $\wpo$ or $\phi_o$ do not have the same values in
all cases but depend on $\Ompi$ as can be seen from
 eqs.(\ref{wo}) and (\ref{po}). However, the dependence on
 $\Ompi$ is mild since for example for $n=1$ one has that $\phi_o$
only depends as $\sqrt{\Ompi}$. For this reason once $H_o$ and
$\Ompi$
are fixed we still have a wide range of values of $\Ompi$ giving
the correct phenomenology, i.e. $\Delta \Ompi =45\%$. For $\Ompi \ll 1$,
i.e. $y_i=c\Lm_c/H_i \ll 1$, with a fixed $\Lm_c$ one has
a larger $H_i$ and the time of expansion up to present day
is also larger. In this case
it is posible that the $\phi$ field has already reached
its tracker value, even for $n<5$, and we would have $\wpo=w_{tr}=-2/(2+n)$.
Indeed, it can be seen from eq.(\ref{po}) that a $\Ompi \ll 1$
renders a $\phi_o \sim n\sqrt{\Ompo/6}$ and
independent of the initial conditions\footnote{We would like to stress
out that eqs.(\ref{po}) and  (\ref{yp}) are valid only
for $0\leq \wp+1 \ll 1$ and in the tracker regime this region
is no longer satisfied.}. For any initial conditions
 we will end up with $-1\leq \wpo \leq
w_{tr}$.

\section{Quintessence Restriction on $n$}\la{sec12}

Before analyzing the quintessence restriction imposed on $n$ we
would like to comment on the value of $c$
 in the potential eq.({\ref{pot}). It can take
 different values with different physical interpretations.

For simplicity of arguments let us assume that we
have no kinetic energy at the beginning and that
$\rho_{\phi i}=V_i=c^2\Lm_c^4$. The initial energy density is
$\Ompi=\rho_{\phi i}/\rho_c$. Taking
$\rho_{\phi i}=c^2\Lm_c^4$, with $c=2\nu$, and $\rho_c=E^4$
we have that the initial energy density
is given by $\Ompi=(2\nu)^2 (\Lm_c/E)^4$. In this
case we see that the initial energy density depends
on the ratio between the condensation scale $\Lm_c$
and the critical energy density $E=\rho_c^{1/4}$.

Another possibility is to take
$\rho_{\phi i}=g_{\phi}T^4$ and $\rho_{c}=g_{Tot}T^4$
where $g_{Tot}, g_{\phi}$ are the total and quintessence
number of degrees of freedom, respectively, at a temperature
$T$. In this case we have $\Ompi=g_{\phi}/g_{Tot}$
and it only
depends  on the ratio of quintessence and total
degrees of freedom and not on the energy scales.
The condensation scale is then $\Lm_c=(g_{\phi}/c^2)^{1/4} T$
and  $\rho_{\phi i}=
g_{\phi}T^4=c^2\Lm_c^4=V_i$.

Let us now,  study the  restrictions to the
values of  $n$. In both cases, mentioned above,
we get similar restrictions so we will only consider
the first one.

For a fixed $n$ we  take the
following conditions: $h_o$ must be in the range $.7 \pm .1$
 and $\Ompo$ must belong to the interval
 $0.6-0.8$. We can restrict $n$ according to different physical
arguments. These restrictions are encountered while solving
numerically the differential system settled by equations
(\ref{eqFRW2}), the first restriction comes from the observational
value of the relevant parameter $\wpo$. The limit $\wpo<-2/3$ can
be translated to $\weff < -0.7$ \ci{w}. Notice that in this
potentials $\weff < \wpo$ contrary to the general arguments
\ci{tracker} and the reason is that $\wp$ is still growing by today.
From this analysis we find that, in order to fulfill the $\weff$
condition, $n$ must be smaller than $2.74$ as shown in Figure
\ref{weff} for $\Ompi=0.25$. The value of $\wpo$ depends not only
on $n$ but also on $\Ompi$ and it decreases with increasing
$\Ompi$. If we fix $n$ and increase $\Ompi$ we find that examples
originally discarded by the $\weff<-0.7$ restriction now enter the
physically permitted group. The example with $n=3$ is depicted in
Figure \ref{weff2}. An equipartition value of $\Ompi$ is 0.25
but if we allow $\Ompi$ to be as large as 0.75 than the
restriction on $n$ is only $n < 5.2$.
\begin{figure}[htp!]
\begin{center}
\includegraphics[width=9cm]{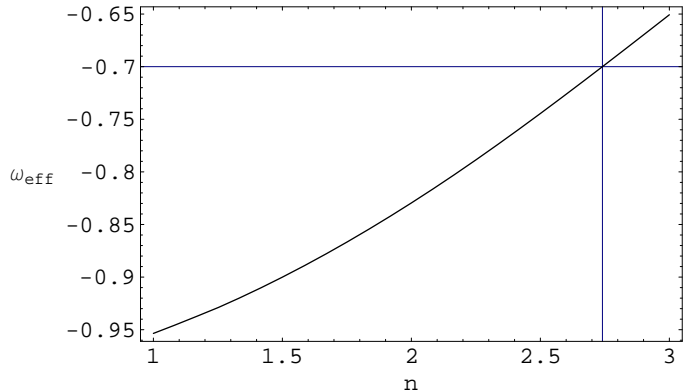}
\end{center}
\caption{\small{Restriction over $n$ from $\weff< -0.7$
 with $c=1$ and $\Ompi=0.35$)}}\la{weff}
\end{figure}
\begin{figure}[htp!]
\begin{center}
\includegraphics[width=9cm]{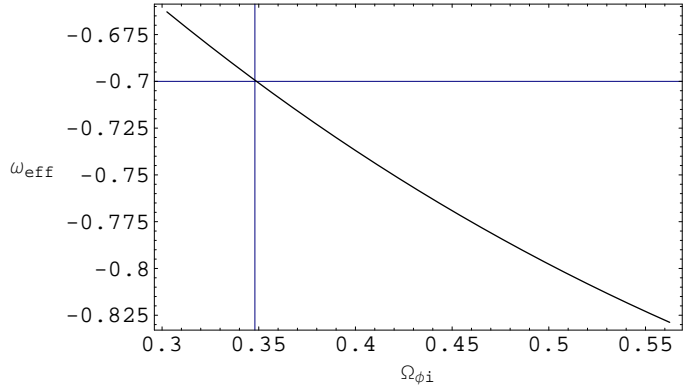}
\end{center}
\caption{\small{Restriction $\weff< -0.7$ avoided by increasing
$\Ompi$ with $n=3$ ($c=1$)}}\la{weff2}
\end{figure}

Other restriction comes from big bang nucleosynthesis (NS) results
that require $\Omp(\mathrm{NS})<0.1$ at the energy scale range of
NS: $0.1-10\ \mathrm{MeV}$ \cite{NS}. To account for this we have
to either consider $\Ompi<0.1$ or, for example, if $\Ompi=0.25$ we
must take out $1.2<n<2.1$ because for this range the initial
value $H_i$ lies within the range of values of $H_{\mathrm{NS}}$.
For all the values of $n$ allowed by $\weff$ and NS restrictions,
a variation of $45 \% $ on $\Ompi$ can be performed without
disturbing the permanence into the observed ranges of $H_o$ and
$\Ompo$.

\section{Unification of Gauge Couplings Constants}\la{unif}

The condensation scale $\Lm_c$ used in eq.(\ref{pot})
is from an elementary particle point of view an arbitrary scale
that sets the energy scale of the phase transition.
However, if   the inverse
power potential eq.(\ref{v}) is obtained from a non-abelian
asymptotically free gauge group then we can relate the
condensation scale $\Lm_c$ to other energy scales using
the renormalization group equation.  The one-loop evolution of the
gauge coupling constant for an $SU(N_c)$ gauge group with $N_f$ chiral fields
gives a condensation scale  (strong coupling $g^{-2}\ll 1$)
 \be \la{lm}
\Lm_c= \Lm_x e^{-\fr{1}{2 b_o g^2_x}}
 \ee
where $b_o=(3 N_c-N_f)/16\pi^2$ is the one-loop beta function and
$\Lm_x, g_x$ are arbitrary energy scale and coupling constant,
respectively, which include high energy scale where the original
chiral fields $Q$ are weakly coupled.

It is well known that the gauge coupling constants of the standard
model get unified at a energy scale
$\Lm_{gut} \simeq 10^{16}\mathrm{GeV}$ with a coupling constant $g_{gut}
\simeq \sqrt{4 \pi/ 25.7}$.

We want to impose to our model gauge coupling unification, i.e.
the coupling of the gauge group responsible for quintessence
should be  unified at $\Lm_{gut}$ with the standard model gauge groups  \ci{chris1}.
In this case we  require $\Lm_x, g_x$ in eq.(\ref{lm}) to take the values
$\Lm_x=\Lm_{gut}, g_x=g_{gut}$ and we
have $\Lm_c = \Lm_{gut}exp[-1/2b_og^2_{gut}]$.
This is not a necessary condition but opens
the possibility of thinking of the model as coming from string
theory after compactifying the extra dimensions or on a Grand
Unification Scheme where all gauge coupling constants are unified.

Of course, not all values of $N_c, N_f$ will give an acceptable
phenomenology. This is because the cosmological evolution
of $\phi$ and the gauge coupling unification set independent
 constrains on the condensation scale $\Lm_c$ and on $N_c, N_f$.
 From a cosmological
 point of view $\Lm_c$ depends on
the inverse power $n$ (see eq.(\ref{lmm})), which is a function of
$N_c, N_f$, and from gauge unification $\Lm_c$ depends also on
$N_c, N_f$ through $b_o$ (see eq.(\ref{lm})). These two constrains
reduce drastically  the allowed  values of $N_c, N_f$ and
we also require $N_f\geq 1$ and $N_c \geq 2$ to be integers.

 In table \ref{constr}
we give the different values of $N_c, N_f, \nu$ for which
we have gauge coupling unification.
We can see that there are only a small number of
possible models (11). The first five models have an $N_f$ which differs
form an integer  by less than $0.05$ while the other
6 models differ at most by $0.10$. All other combinations of $N_c, N_f, \nu$
have a larger discrepancy and do not lead to $\Lm_c=\Lm_u$.
If we further constrain  the models to agree with the
cosmological observations (i.e. $\wpo < -0.7$ requiring $n<5$)
we are left with only 4 models (number
1,2,3,11 of table \ref{constr}). All of these 4 models
have $n<2$ and the quantum corrections to the Kahler potential
are, therefore, not dangerous. Notice as well that only two models
(4,9) have $\nu=N_f$ and in both cases $n>12$.
\begin{table}
\begin{center}
\begin{tabular}{|c|c|c|c|c|c|}
\hline
 Num& $ N_c$ &$N_f$ &$ \nu$ & $n$ &$ \Lambda_c $(GeV)\\ \hline
\hline
 1& 3 & 5.98 & 1 & 0.66 & $6\times 10^{-8}$ \\
 2 &6 & 14.97 & 3 & 0.66 & $6.9 \times 10^{-8}$ \\
  3& 7 & 18.05 & 4 & 0.55 & $1.6 \times 10^{-8} $\\
 4& 8 & 5.97 & 5.97 & 13.83 &$ 1.3 \times 10^{12}$ \\
 5&  8 & 6.96 & 3 & 13.55 & $7.6 \times 10^{11}$ \\ \hline \hline
 6& 3 & 1.90 & 1 & 5.66 & $1.3 \times 10^{6} $ \\
  7& 5 & 3.91 & 2 & 9.38 & $ 4.7\times 10^{9} $ \\
  8& 6 & 5.09 & 2 & 10.85 &$ 3.7\times 10^{10} $ \\
 9& 7 & 5.08 & 5.08 & 12.64 &$ 3.9\times 10^{11} $ \\
  10& 8 & 20.90 & 4 & 6.75 & $2.4  \times 10^{-7} $ \\
  11& 8 & 21.10 & 5 & 0.47 & $5.7  \times 10^{-9} $ \\
  \hline
\end{tabular}
\end{center}
\caption{\small{Models with $\Lm_c=\Lm_u$. The first 5 have $\Delta N_f <0.05$
while the other 6 have $\Delta N_f < 0.10$ discrepancy from an integer}}
\la{constr}\end{table}
\begin{table}[p!]
\begin{center}
\begin{tabular}{|l|l|l|l|l|l|l|}
\hline $n$&$H_i$ (GeV)&$\rho_{\phi i}^{1/4}$(GeV)&
$\Lm_c$ (GeV) & $\wpo$& $\weff$&$N_{total}$\\ \hline\hline
$1/2$&$1.33\times
10^{-35}$&$7.48\times10^{-9}$&$3.74\times10^{-9}$&$-0.97$&$-0.98$&10.72\\
$2/3$&$1.16\times 10^{-33}$&$6.99\times10^{-8}$ &
$3.49\times10^{-8}$&$-0.97$&$-0.98$&12.96\\ $1$&$3.46\times
10^{-30}$&$3.82\times10^{-6}$&$1.91\times10^{-6}$&$-0.93$&$
-0.95$&16.97\\ $2$&$4.68\times
10^{-22}$&$4.44\times10^{-2}$&$2.22\times10^{-2}$&$-0.77$&$
-0.83$&26.33\\ $18/7$&$1.68\times
10^{-18}$&$2.66$&$1.33$&$-0.65$&$ -0.73$&30.42\\ $3$&$3.41\times
10^{-16}$&$37.92$&$18.77$&$-0.57$&$ -0.65$&33.07\\ \hline
\end{tabular}
\end{center}
\caption{\small{Numerical solutions for different values of $n$ with
$c=1$ and $\Ompi=0.25$. }}\label{table1}
\end{table}

\begin{table}[p!]
\begin{center}
\begin{tabular}{|l|l|l|l|l|l|l|}
\hline $\Ompi$&$H_i$ (GeV)&$\rho_{\phi i}^{1/4}$ & $\Lm_c$
(GeV) &$\wpo$& $\weff$&$N_{total}$\\ \hline\hline $1\times
10^{-5}$&$5.48\times 10^{-14}$&$480.90$&$19.12$&$-0.55$&$
-0.50$&35.69\\ $0.25$&$3.34 \times
10^{-16}$&$37.54$&$18.77$&$-0.57$&$ -0.65$&33.07\\
$0.5$&$2.85\times 10^{-16}$&$34.68$&$20.62$&$-0.76$&$
-0.83$&32.88\\ $0.75$&$3.05\times
10^{-16}$&$35.88$&$23.61$&$-0.88$&$ -0.92$&32.75\\
$0.99$&$5.13\times 10^{-16}$&$46.53$&$32.82$&$-0.97$&
$-0.98$&32.20\\ \hline
\end{tabular}
\end{center}
\caption{\small{Numerical solutions for different values of $\Ompi$
with $c=1$ and $n=3$. }}\label{table2}
\end{table}
As an example of a model with gauge coupling unification we have a gauge
group $N_c=3$ with $N_f=6$ and $\nu=1$
which entails a value $n=2/3$ according to $n=2+\fr{4
\nu}{N_c-N_f}=2/3$ \ci{chris1}. For this model we find, from the
numerical solution, a total time $N_{total}=12.96$ which does not
superposes on the NS range $19.6-27.2$. The $\weff < -0.7$
restriction is satisfied with $\weff=-0.98$ and the condition from
experimental central values $h_o=0.7   $ and
$\Ompo=0.7$ is also fulfilled taking $\Ompi=0.25$. The condensation scale is
$\Lm_c=4.2\times10^{-8}\ \mathrm{GeV}$. A full analysis of this
model is presented in \ci{ax.unif}.

\section{Further Examples}\la{examp}

We have already given some example in the previous sections. In
table \ref{table1} and \ref{table2} we show the numerical results
for different values of $n$ with initial condition $\Ompi=0.25$
and for different initial condition with n=3 fixed, respectively.

Other interesting examples are when $\nu=N_f$. The condition $n<2.74$
(for $\Ompi \leq 0.25$) requires $N_f/N_c <0.15$ and therefore $N_c>7$.
For $N_c=8, N_f=1$ one has $n=18/7 \simeq 2.57$ and using $\Ompi=0.25$
and one obtains $\wpo=-0.65,\weff= -0.73, \phi_o=2.01, N_o=30.3$ and $M_i=2.3$ GeV
$ \Lm_c=1.6$ GeV  while
for $N_c=7, N_f=1$ one has $n=8/3, \wpo=-0.64, \weff=-0.71,
\phi_o=2.04, N_o=30.9, M_i=4.2 GeV, \Lm_c=3 GeV$.
However, these models do
not have $\Lm_c=\Lm_u$, i.e. they are not unified with the SM gauge
groups.

\section{Conclusions}\la{concl}

We  studied negative power potentials and we constrain the
initial conditions an the power of the potential to satisfy
the SN1a results. For $n$ larger than 5, the scalar field $\phi$
has already  reached its tracker value and $\wpo$ is too large.
So, we need to concentrate on potentials with $n < 5$ to comply with
SN1a results. We gave a semi analytic solution to $\wpo$ and
$\Lm_c$ in terms of $H_o, \Ompi$ and $n$ and we have solved numerically for
some  relevant cases. We obtained that $\wpo$
depends on $\Ompi, n$ and it decreases with increasing $\Ompi$
while it becomes smaller for larger $n$. If we assume equipartition
initial conditions with $\Ompi \leq 0.25$ than $n$ is constrained
to be smaller than $n<2.74$, however, if we allow for $\Ompi=0.75$
the constraint is relaxed to $n<5.2$. We have shown that one can
vary the initial conditions up to $45\%$ without spoiling
the observational cosmological values at present time.  For any
 initial conditions we will end up with $-1\leq \wpo \leq w_{tr}$.

We have seen that the negative power potentials can be derived
from Affleck's potential and in order to avoid problems with the
Kahler potential one requires $n<2$ which implies that $N_f>N_c$
and that not all condensate become dynamically (i.e. $\nu \neq
N_f$). For $\nu=N_f$ one needs $N_f/N_c < 0.15$ to have $\wpo<-2/3$.
 Furthermore, we have shown that it is
possible to have a quintessence model with
gauge coupling unification for all gauge groups, Standard
Model and the gauge group responsible for quintessence, but the
number of models is quite limited (4).

This work was supported in part by CONACYT project 32415-E and
DGAPA, UNAM project IN-110200.

  \thebibliography{} \footnotesize{
\bib{SN1a} {A.G. Riess {\it et al.}, Astron. J. 116 (1998) 1009; S.
Perlmutter {\it et al}, ApJ 517 (1999) 565; P.M. Garnavich {\it et
al}, Ap.J 509 (1998) 74.}

\bib{CMBR} {P. de Bernardis {\it et al}. Nature, (London) 404, (2000)
955, S. Hannany {\it et al}.,Astrophys.J.545 (2000) L1-L4}

\bib{w}{S. Perlmutter, M. Turner and  M. J. White,
Phys.Rev.Lett.83:670-673, 1999; T. Saini, S. Raychaudhury, V. Sahni
and  A.A. Starobinsky, Phys.Rev.Lett.85:1162-1165,2000 }

\bib{tracker} I. Zlatev, L. Wang and P.J. Steinhardt, Phys. Rev.
Lett.82 (1999) 8960;  Phys. Rev. D59 (1999)123504

\bib{1/q} {P.J.E. Peebles and B. Ratra, ApJ 325 (1988) L17; Phys. Rev. D37 (1988) 3406}

\bib{2/q}{J.P. Uzan, Phys.Rev.D59:123510,1999}

\bib{bine}{P. Binetruy, Phys.Rev. D60 (1999) 063502, Int. J.Theor.
Phys.39 (2000) 1859}

\bib{mas} A. Masiero, M. Pietroni and F. Rosati, Phys.
Rev. D61 (2000) 023509}

\bib{chris1}{A. de la Macorra and C. Stephan-Otto, Phys. Rev. Lett. 87,
 (2001) 271301, astro-ph/0106316}

\bib{generic}  A.R. Liddle and R.J. Scherrer, Phys.Rev.
D59,  (1999)023509

\bib{mio.scalar}{A. de la Macorra and G. Piccinelli, Phys.
Rev.D61 (2000) 123503}

\bib{liddle}E.J. Copeland, A. Liddle and D. Wands, Phys. Rev. D57 (1998) 4686

\bib{r=m}{E.Kolb and M.S Turner,The Early Universe, Edit. Addison Wesley 1990}

\bib{NS} {K. Freese, F.C. Adams, J.A. Frieman and E. Mottola, Nucl. Phys. B 287
(1987) 797; M. Birkel and S. Sarkar, Astropart. Phys. 6 (1997)
197.}

\bib{Affleck}{I. Affleck, M. Dine and N. Seiberg, Nucl. Phys.B256
(1985) 557}

\bib{ax.asy}{C.P. Burgess, A. de la Macorra, I. Maksymyk and F. Quevedo
Phys.Lett.B410 (1997) 181}

\bib{ax.coinci}{A. de la Macorra, Int.J.Mod.Phys.D9 (2000) 661 }

\bib{unif}{U. Amaldi, W. de Boer and H. Furstenau, Phys. Lett.B260
(1991) 447, P.Langacker and M. Luo, Phys. Rev.D44 (1991) 817}

\bib{ax.unif} A. de la Macorra hep-ph/0111292

}
\end{document}